\documentclass[12pt,preprint]{aastex}
\usepackage{natbib}

\def\spit{{\it Spitzer}}
\def\70um{70 \micron}
\def\24um{24 \micron}
\def\um{\micron}
\def\ld{L_{\rm dust}/L_{\star}}

\def\Mstar{M_{\star}}

\def\MEarth{M_\oplus}

\def\Mp{M_{\rm p}}
\def\MJup{M_{\rm Jup}}

\def\gapp{\lower 3pt\hbox{${\buildrel > \over \sim}$}\ }
\def\lapp{\lower 3pt\hbox{${\buildrel < \over \sim}$}\ }
\def\proptosim{\lower 3pt\hbox{${\buildrel \propto \over \sim}$}\ }

\def\arcsec{$^{\prime\prime}$}

\begin{document}

\title{Planets and IR Excesses: \\
Preliminary Results from a \spit/MIPS Survey
of Solar-Type Stars}

\author{
C. A. Beichman$^{1}$, G. Bryden$^{1}$, 
G. H. Rieke$^{2}$, J. A. Stansberry$^{2}$, D. E. Trilling$^{2}$,
K. R. Stapelfeldt$^{1}$, M. W. Werner$^{1}$, 
C. W. Engelbracht$^{2}$, 
M. Blaylock$^{2}$, K. D. Gordon$^{2}$, C. H. Chen$^{1}$, 
K. Y. L. Su$^{2}$, \& D. C. Hines$^{2}$}
\affil{1) Jet Propulsion Lab, 4800 Oak Grove Dr, Pasadena, CA 91109} 
\affil{2) Steward Observatory, University of Arizona, 933 North Cherry Ave, Tucson, AZ 85721}

\begin{abstract}
As part of a large \spit/MIPS GTO program, we have searched for infrared
excesses due to debris disks toward 26 FGK field stars known from
radial velocity (RV) studies to have one or more planets.  
While none of these stars show excesses at \24um, we have detected \70um
excesses around 6 stars at the 3-$\sigma$ confidence level.  
The excesses are produced by cool material ($<100$ K) located
beyond 10 AU, well outside the ``habitable zones'' of these systems and
consistent with the presence of Kuiper Belt analogues with $\sim$100 times
more emitting surface area than in our own planetary system.  
These planet-bearing stars are, by selection for RV studies, typically
older than 1 Gyr, and the stars identified here with excesses have a
median age of 4 Gyr.
We find a preliminary correlation of both the frequency and
the magnitude of dust emission with the presence of known planets. 
These are the first stars outside the solar system identified as having
both well-confirmed planetary systems and well-confirmed IR excesses. 

\end{abstract}


\section{Introduction}

First discovered by the Infrared Astronomical Satellite (IRAS),
the ``Vega phenomenon'' is characterized by a significant deviation from
the expected Rayleigh-Jeans emission of quiescent main-sequence stars 
\citep{Aumann90, Gillett86}.
The excess IR emission
from other stars has been attributed to dust orbiting 
at distances anywhere from less than 1 to more than 100 AU,
analogous to the zodiacal cloud and the Kuiper belt in our own solar system.
Due to effects such as radiation pressure and Poynting-Robertson drag,
the lifetime of the dust is generally much shorter than
the age of the system; 
any dust observed must have been recently produced.
In the solar system, dust is generated by collisions between larger
bodies in the asteroid and Kuiper belts, as well as from outgassing
comets. Extrasolar systems with IR excess presumably have their own supply
of large, solid planetesimals and, perhaps, large planets like those
in our system.

IRAS found that approximately 15\% of main-sequence stars
\citep{Plets99}, predominantly A and F stars, show excesses
attributable to dust in orbit around the stars at distances between a
few and a few hundred AU and heated by absorbed starlight to
temperatures of  30-300 K \citep{Backman93}. 
Fractional luminosities, $\ld$, are observed as large as $10^{-3}$ for
stars with bright excesses such as $\beta$ Pic and as small as
$10^{-5}$ for stars at the detection threshold.
These values can be compared to our solar
system: $10^{-8}$ to $10^{-7}$ for the hot material in the zodiacal cloud
\citep[1-3 AU;][]{Dermott02}
and $10^{-7}$ to $10^{-6}$ predicted for the cooler Kuiper
Belt region beyond 10 AU \citep{Stern96}, which is constrained to be
$<10^{-5}$ by IRAS \citep{Aumann90,Teplitz99}. 
Observations by the Infrared Space  
Observatory \citep[ISO;][]{Habing01,Spangler01} suggest a decline in
the fraction of stars with excess IR emission with time, but with the
possibility of finding modest ($\ld >10^{-5}$) excesses among older
stars \citep{Decin00, Decin03}. \spit\ and IRAS results toward more
than 200 A stars confirm a general decline in the average amount of
emission with little \24um emission seen for sources older than 100
Myr but with large variations within each age group studied
\citep{Rieke05}. 
The variations are probably due in part to sporadic replenishment of
dust clouds by individual collisions between large, solid bodies, but
also are likely to reflect a range in mass and extent for the initial
planetesimal disk. 

Images in the optical and near-IR \citep{Heap00,Schneider99},
submillimeter \citep{Holland98,Greaves98,Holland03} and from
\spit\ itself \citep{Stapelfeldt04} have revealed that in some cases
the dust exists in flattened disks with warps, gaps, and blob-like
structures that have been attributed to the gravitational influence of
planets \citep{Wyatt02,Dermott02}. 
Still, while the frequency and magnitude of IR excess are strongly
correlated with stellar youth, no such correlation has yet been found
between IR excess and the presence of planets.
Even before the first confirmed discovery of an extrasolar 
planet orbiting a Sun-like star \citep[51 Peg b;][]{Mayor95,Marcy95},
attempts had been made to find links between planets and 
dusty debris around other stars \citep[e.g.\ $\epsilon$ Eri;][]{Backman93}.
While early radial-velocity searches for planets 
found a planet-like trend for $\epsilon$ Eridani \citep{campbell88}, 
the candidate planet remains controversial even today due to the
extremely high level of chromospheric activity on this young star. 
The 7-year periodicity found in the star's radial velocity
\citep{campbell88,Cumming99,Hatzes00} may be induced by a planet
orbiting at several AU \citep{Hatzes00} or may instead be a reflection
of the star's poorly understood magnetic cycle.
Conversely, 55 Cnc has a well-known planetary system, but claims of a
dust disk based on visible imaging \citep{trilling00}
and ISO photometry \citep{dominik98}
have been ruled out by HST coronography \citep{Schneider01},
submillimeter mapping \citep{RayJay02},
and \spit/MIPS measurements presented here.
In spite of the many studies specifically designed to find
a connection between dust and planets,
no systems with both well-confirmed planets and 
well-confirmed IR excess have yet been observed.
In fact, the data to date appear to indicate an anti-correlation 
between dust and planets \citep{Greaves04}.

The longstanding assumption that stars besides the Sun might harbor
both planets and extended dust emission has still not been verified.
The aim of this paper is to use 
the Multiband Imaging Photometer 
on the {\it Spitzer Space Telescope} \citep[MIPS;][]{Rieke04}
to directly address this outstanding issue by searching for excess IR
emission from known planet-bearing stars. 
MIPS provides unprecedented sensitivity at mid-IR wavelengths and as
such is an ideal instrument for searching for the dust emission from
solar-type stars. 
While the focus of this paper is on observations of stars which are
known to have planets, for purposes of noise calibration and to greatly
improve our statistical analysis, the results for planet-bearing stars
are considered within the context of a larger program - 
the Volume Limited Survey (VLS).
VLS is a \spit\ GTO program designed to search for excesses around a
broad sample of 155 nearby, F5-K5 main-sequence field stars, 
sampling wavelengths from 8-40 \um \ with IRS and 24 \& \70um
with MIPS.  
This survey consists of two components - stars with planets and those
without.
The stars without known planets represent a true volume-limited sample
selected for expected \70um brightness and low levels of infrared cirrus.  
The sample includes G and late F stars out
to 25 pc and early K stars out to 15 pc.  
This component of VLS consists of 114 stars (without planets), 58 of
which have already been observed.
These stars will be discussed in more detail in a separate paper
\citep{bryden05};
here we focus on the second component of the VLS program -
planet-bearing stars.
The planet-bearing sample was selected from a group of
stars of similar spectral types known to have RV
planets. 
Of 41 planned targets, 26 have been observed to date.
These stars are not necessarily located in 
regions of low infrared cirrus and are typically farther away 
($\sim$15-50 pc) than the non-planet VLS sample.
It is worth noting that the disk around $\epsilon$
Eri, a K0 star located 3.3 pc away, would have been undetectable by
IRAS if that star were located even 10 pc away; its 60
\um \ flux density would then have dropped below the completeness
limit of the Faint Source Catalog \citep{Moshir90}. 
The sensitivity of \spit/MIPS allows the search for disks not just
around the nearest FGK stars, but also around those located at
distances of 10-50 pc around which planets have been identified.  

We present our MIPS observation in \S\ref{observations}, along with 
a full error analysis to determine whether any measured excesses are
statistically significant.
For the systems with IR excess, \S\ref{parameters} attempts to find
a correlation with system parameters such as stellar metallicity/age
or planet mass/semi-major axis.
Before the concluding remarks (\S\ref{conclusions}),
we describe how our MIPS observations constrain the dust properties
in each system (\S\ref{ldsec}).

\section{Observations}\label{observations}

During its first year of operation \spit\ observed 
from the VLS sample 26 stars with RV planets (Table \ref{planetparam}) 
and an additional 58 stars without any known planets. 
All stars were observed at \70um.
One planet-bearing star, HD128311, was not observed at
\24um due to scheduling constraints, but its spectrum at this
wavelength was measured with IRS Long-Low and was found to
be consistent with no excess shortward of 30 \um. 
The remainder of the planet-bearing stars were observed at \24um and 
detected with high signal-to-noise ratio (SNR).

\subsection{Data Reduction}

Our data reduction is based on the DAT software developed by the MIPS
instrument team \citep{Gordon04a}. 
Because the data pipeline is still under development and the adopted
calibration numbers are likely to change with further
improvements in the overall MIPS data analysis, we list our chosen set
of data reduction parameters in detail. 

For each of the \24um and \70um bands, the instrument calibration was
derived from $\sim$75 observations of $\sim$35 different stars.  
To minimize systematic uncertainties, a mix of stellar types was used:
most of the calibrators are K giants, several are A stars, and a few
are solar analogs.  The \24um photosphere predictions were derived
from stellar templates \citep[e.g.][]{Cohen99} for K giants, 
from Kurucz models \citep[e.g.][]{Kurucz03} for A stars, and
from the solar spectrum \citep[e.g.][]{tobiska00} for the solar analogs.  
The \70um photospheric predictions were obtained similarly, except
that the analytic spectral functions of \citet{Engelke92} were used to
extrapolate the K giant templates out to \70um.

At \24um, we carried out aperture photometry on reduced images using an
aperture 6 camera pixels (1 pixel=2.5\arcsec \ at \24um) 
in radius, a background annulus from 12 to 17 pixels,
and an aperture correction of 1.15.  
The flux level is calibrated at 1.047 $\mu$Jy/arcsec$^2$/(DN/s), 
with a default color correction of 1.00 appropriate for sources warmer
than 4000 K at a weighted-average wavelength of 23.68 \um.  
Final images were mosaiced from individual frames with quarter-pixel
subsampling.

At \70um we used images processed beyond the standard DAT software in
order to correct for time-dependent transients, corrections
which can significantly improve the sensitivity of the measurements
\citep{Gordon04b}.
At \24um, a large enough aperture was chosen so as to integrate all of 
the flux out to roughly the first Airy ring. 
Because the accuracy of the \70um data is limited by background noise,
rather than instrumental effects, such a large aperture contains an
undesirable amount of sky fluctuations.
In order to maximize signal-to-noise, a smaller aperture is more
appropriate at \70um - just 1.5 pixels in radius.
With a 4 to 8 pixel radius sky annulus, this aperture size requires a
relatively high aperture correction of 1.79. 
The flux level is calibrated at 15,800 $\mu$Jy/arcsec$^2$/MIPS\_70\_unit, 
again with a default color correction of 1.00
(MIPS\_70\_units are based on the ratio of the measured signal to the
stimulator flash signal). 
Images were mosaiced from individual frames with half-pixel
subsampling.
For both the \24um and \70um data, neighboring point sources were
subtracted from the images before measuring the sky brightness.
Also for both wavelengths, we fix each target's position, rather than 
centroid onto the peak of the local flux distribution.
If the target position is not fixed, the centroid position tends to
float toward any local noise fluctuations, resulting in an
inappropriately high flux measurement, particularly for noisy fields.
The fact that no centering of the aperture is required is a 
reflection of the excellent telescope pointing intrinsic to \spit\
\citep[$<$1\arcsec;][]{werner04}.

\begin{figure}
\begin{center}
\includegraphics[width=4.7in,angle=-90]{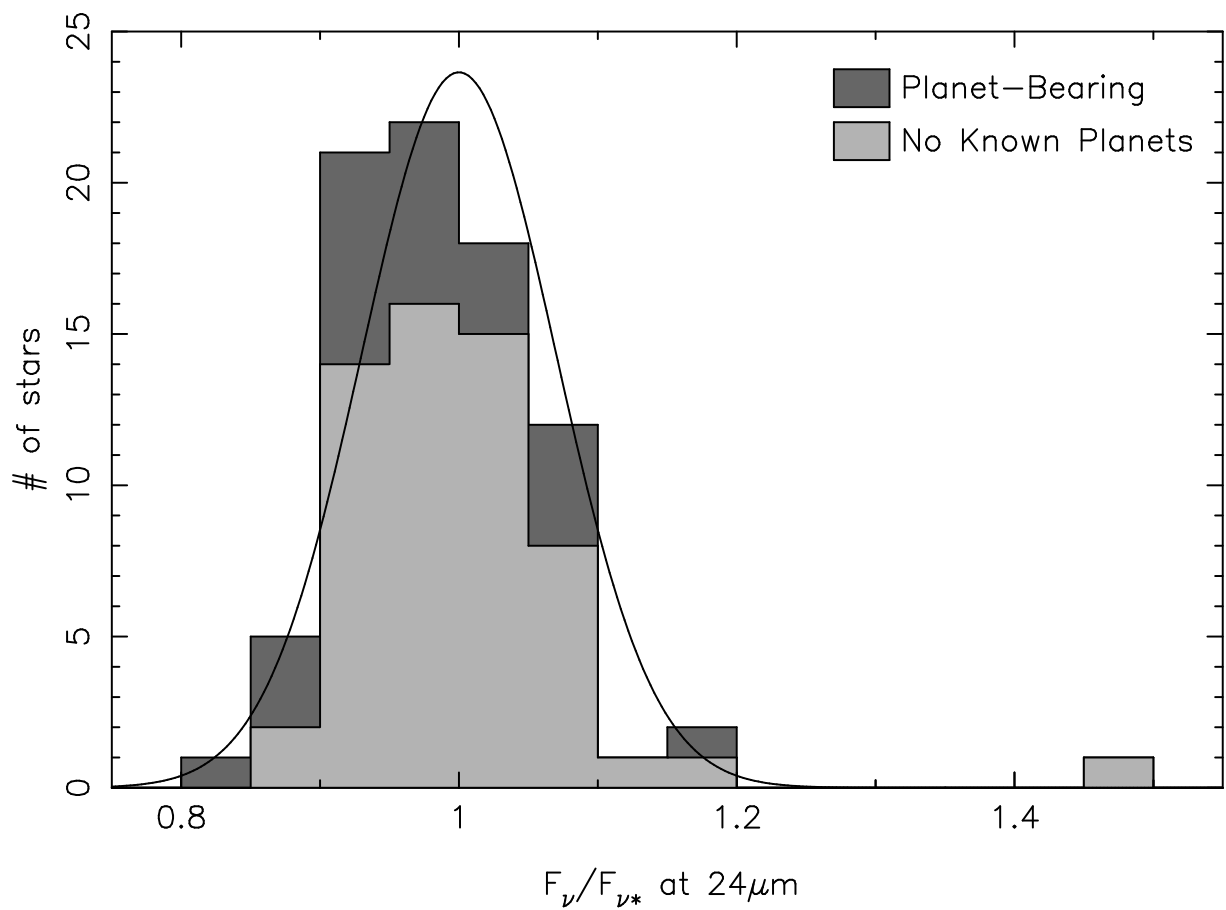} 
\end{center}
\caption{Distribution of \24um fluxes relative to the expected
photospheric values for a sample of 84 stars.  A gaussian distribution
with dispersion 0.07 is shown for comparison. One star
\citep[HD69830;][]{Beichman05} clearly stands out from the main
population of stars with no excess emission above their stellar
photospheres. In this histogram and those that follow, stars with
planets are shaded with dark grey while stars without known planets
are designated by light grey.
\label{f24k}}
\end{figure}

In order to determine whether any of our target stars have an IR excess,
we compare the measured photometry to predicted photospheric levels. For
convenience in extrapolating the photospheric emission we have fitted
Kurucz models \citep{Castelli03,Kurucz03} to a combination of short
wavelength data taken primarily from the Tycho/Hipparcos (visible),
Johnson (bright near-IR) or 2MASS (faint near-IR), and IRAS (12 \um)
catalogs. After excluding one outlying point for a star with an excess 
\citep[HD69830, which will be discussed elsewhere;][]{Beichman05}, the 83
flux measurements at \24um have an average $F_{\rm MIPS}/F_{\rm
photosphere}$ of 0.99 which, although gratifyingly close
to unity, is not surprising given that the present \spit\ calibration
is based on similar stellar models. 
More importantly for our purposes, the dispersion of 
$F_{\rm MIPS}/F_{\rm photosphere}$ is 0.07 for this sample
(Fig.~\ref{f24k}). 

At \70um, 73 out of 84 stars (both planet-bearing and without planets) 
are detected with signal-to-noise ratio $>3$. This is in contrast with
previous IR surveys of AFGK stars with ISO, in which only half of the
stars without excess were detected \citep{Habing01}. While the
sensitivity of these \spit\ observations is roughly  a factor of 10
better than previous data, \spit's sensitivity is limited by
extragalactic source confusion and cirrus (see below) which will make
it difficult to look for excesses around stars much fainter than those
discussed here. 

\begin{figure}
\begin{center}
\includegraphics[width=4.7in,angle=-90]{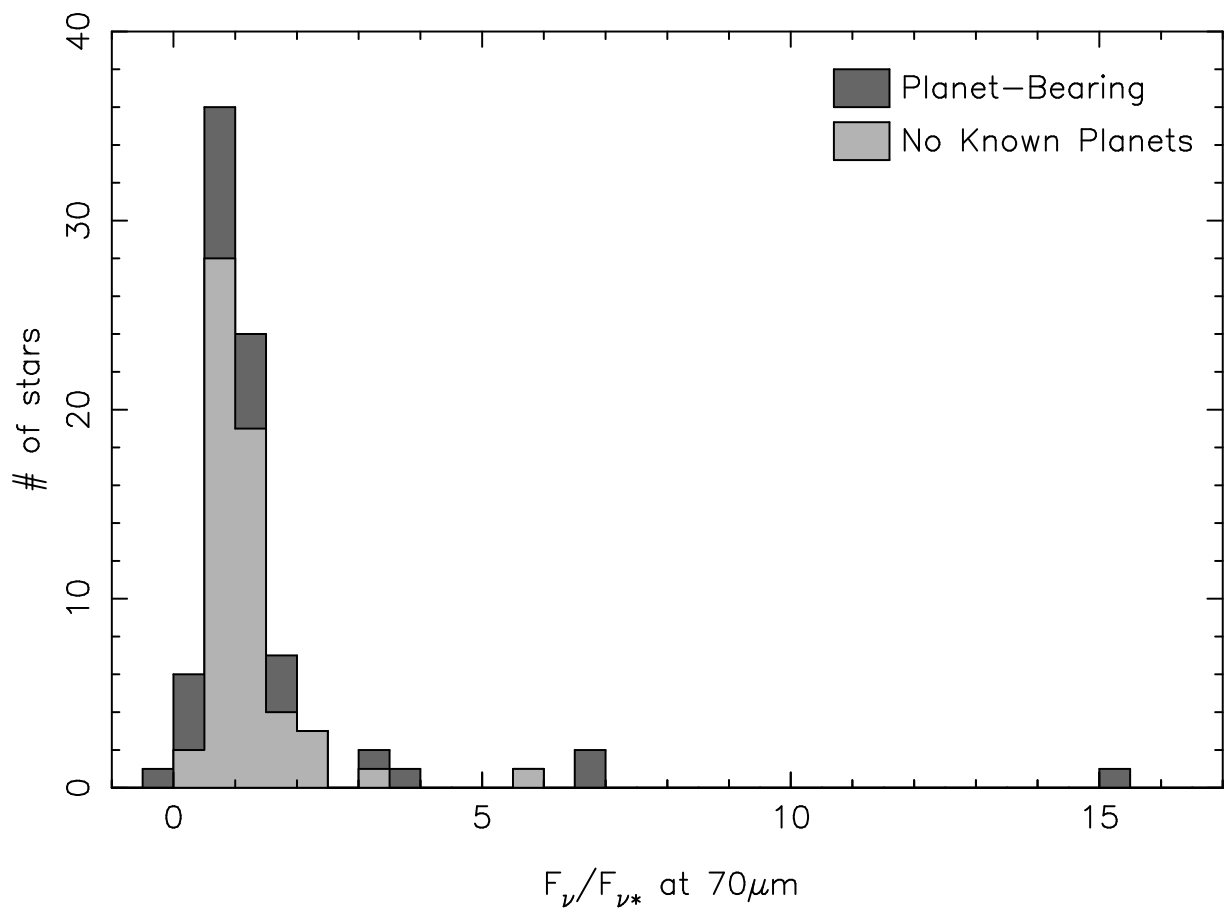} 
\end{center}
\caption{Distribution of \70um fluxes relative to the expected photospheric values for a sample of 84 stars.  While most stars cluster around 1, where 
their flux is photospheric, several stars show a high degree of excess emission.
Stars with planets are shaded as dark grey
while stars without known planets are designated with light grey.
Although planet-bearing stars make up less than a third of the sample,
four out of the five stars with the highest factor of excess \70um emission
are known to have planets.
\label{f70k}}
\end{figure}

The distribution of \70um flux densities relative to the expected
photospheric values is shown in Fig.~\ref{f70k}.
Unlike the tight distribution of flux ratios at \24um,
several stars have \70um flux much higher than expected from the
stellar photosphere alone.   In one case, the \70um flux is high 
by a factor of 15. Many of the stars (12) will be 
identified in the following as having statistically significant IR excess.
Ignoring these stars with excesses and those with SNR $<3$,
the average ratio of MIPS flux to predicted photosphere is $|F_{\rm
MIPS}/F_{\rm photosphere}| =1.03 \pm 0.04$, consistent with
the present calibration. 
More important than the absolute calibration of $F_{\rm
MIPS}/F_{\rm photosphere}$ is its dispersion, which indicates the 
overall accuracy of the flux measurements and gives an immediate
sense of the minimum level of excess necessary for detection.
The dispersion in the \70um sample is 28\% (excluding the stars with
excesses), considerably higher 
than that in the \24um data (7\%).
The following section (\S\ref{noisesec}) discusses in more detail the
noise levels within the \70um data.

\begin{figure}
\begin{center}
\includegraphics[width=4.8in]{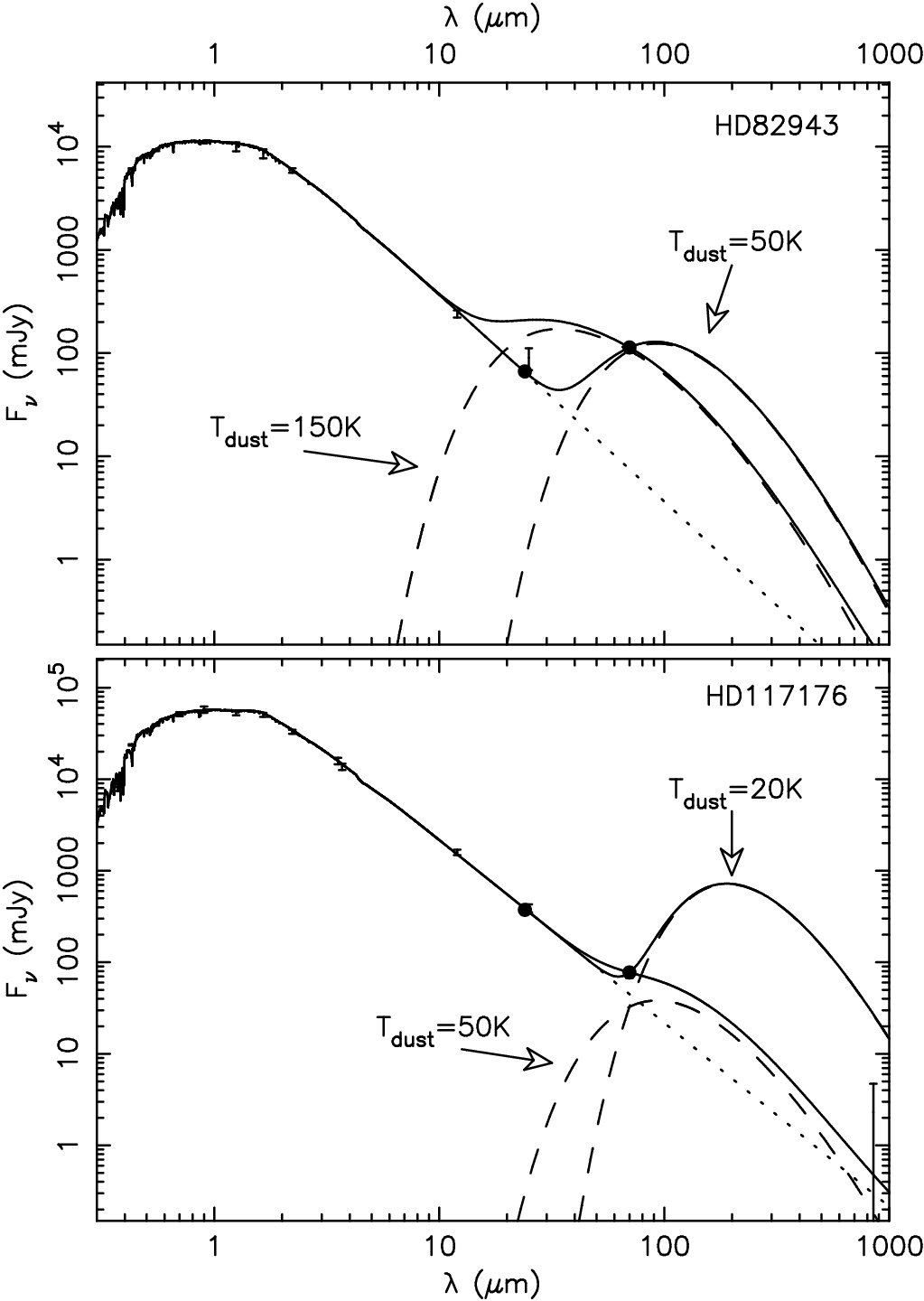} 
\end{center}
\caption{Spectral energy distributions for HD82943 and HD117176.
In addition to our 24 and \70um \spit\ data (dark circles), we also show
optical measurements and IRAS fluxes at 12 and 25 \um.
For HD117176, a sub-mm (850 \micron) constraint is also available
\citep{Greaves04}. 
The emission from dust at a given temperature
(dashed lines) is added to the stellar Kurucz model (dotted line)
in order to fit the observed excess emission at \70um.
In each plot, two separate fits (two different dust temperatures) are
considered. 
In the case of HD82943 hot dust (150 K) is ruled out by the \24um
observations, while for HD117176 cold dust (20 K) is excluded by the
sub-mm upper limit (see \S\ref{ldsec}
for discussion of the limits on dust temperature and luminosity).
\label{sedboth}}
\end{figure}

Fig.~\ref{sedboth} shows the spectral energy distributions
for two of the planet-bearing stars, HD82943 and HD117176, that have
excess \70um emission. 
Published photometric fluxes for these stars from visible to infrared
are well fit by Kurucz stellar atmospheres 
\citep[dotted lines;][]{Castelli03,Kurucz03}.
The \spit/MIPS \24um fluxes are also well fit by the model atmospheres
(within 1\% for HD82943, 5\% low for HD117176).
The \70um emission, meanwhile, is well above that expected from the 
stellar photosphere alone; an additional component of emission due to
dust must be added to the Kurucz model in order to fit the \70um data.
Because there is only a single \70um measurement of
IR excess, however, the SED can be fit by a range of dust temperatures
and luminosities. 
Constraints on the dust in these systems are discussed in \S\ref{ldsec}.

\subsection{Detection of Excess at \70um}\label{noisesec}

A potential correlation of planets with IR excess is already apparent
in Fig.~\ref{f70k}, in which 4 of the 5 largest \70um flux ratios are for
stars which harbor planets, even though planet-bearing stars make up
less than one-third of the sample. Still, without a proper analysis of the noise levels in each field it is impossible to assess whether the IR excesses are statistically significant.

In general, the sources of noise can be divided into pixel-to-pixel
noise (which we refer to as ``sky'' noise) and errors for the field as a
whole (``systematic'' noise). The pixel-to-pixel noise is a combination of detector/photon noise along with real sky background fluctuations.
For the \24um measurements, photon noise is negligible.
Even with the minimum allowed integration time  (1 cycle of 3 
sec exposures = 48 sec), the sensitivity of MIPS is overwhelming;
even our dimmest source could theoretically be detected in just 
a few milliseconds. Further, the background noise is low: 
galactic cirrus is weak at this wavelength;
the zodiacal emission is relatively smooth across the field of view;
and the confusion limit for distant extragalactic sources is just
0.056 mJy \citep{Dole04}. 

Instead, for the \24um measurements, systematic errors dominate.  
The instrumental contribution to these errors is thought to be
very low: 24 \um \ observations of calibrator stars are stable with
1\% rms deviations over several months of observations
\citep{Rieke04}. 
In addition to any residual instrumental problems, the dispersion in $F_{\rm
MIPS}/F_{\rm photosphere} $ can include errors in the photosphere
extrapolation and the effects of source variability. The fitting of the
photosphere can be as precise as 2\% when the best 2MASS K$_s$
photometry is available \citep{skrutskie00}, but for stars brighter
than K$_s<6$ mag, 2MASS data are less accurate and lower precision
near-IR data and/or shorter wavelength observations must be used. 
Extrapolation from visible data places considerably more weight
on the photospheric models and may result in greater uncertainty in
the predicted photospheric levels at \24um and \70um. 

While the relative contributions of instrumental and fitting errors to
the \24um dispersion are unknown, it is clear that the photometry is
currently limited to 7\% accuracy due to their combined effects. 
While this may improve with further instrument analysis and better 
photospheric models, 3-$\sigma$ detections of excess at \24um
now require measured fluxes at least 20\% above the stellar photosphere.  
Only one of our sources (HD69830) meets this
requirement, with a strongly significant 45\% excess.  The relatively
hot dust emission in this system, which is not yet known to have any
planets, will be discussed in a separate paper \citep{Beichman05}. 

At \70um we planned the number of observing cycles (each with 126 sec
integration time) so that the expected detector/photon noise would be 
$\sim$0.1 of the stellar flux, or equivalently, to achieve a 
signal-to-noise (SNR) $\sim$ 10 on the stellar photosphere. 
Integration times range from a few minutes up to 1 hour.
Within each reduced image, we estimate the sky noise by 
directly measuring the
standard deviation in the background flux when convolved with our
chosen aperture size.  Based on this measured sky noise, we find a
median SNR of $\sim$6, excluding the sources identified as having an excess.
This noise estimate, however, includes contributions beyond just
detector/photon noise.  Unlike at \24um, the fluctuations in the
background at \70um can be significant.  This background, a
combination of galactic cirrus and extragalactic confusion, creates a
noise floor for each field which cannot be improved with increased
integration time.  To try to minimize this problem, the 
target stars in the Volume Limited Sample came from areas of low galactic
 cirrus, as estimated from the IRAS Sky Survey Atlas \citep{ipac94}.
The confusion limit for extragalactic background sources, however, is
unavoidable and sets a strict lower limit for the sky noise at \70um.
\citet{Dole04} find a 5-$\sigma$ confusion limit of 3.2 mJy 
based on source counts from \spit\ \citep{Dole04a}
extrapolated to fainter fluxes with the model of \citet{lagache04}.
In our sample, the lowest (1-$\sigma$) 
noise levels observed toward stars located in clean portions of the
sky are $\sim$2 mJy, somewhat worse than Dole et al.'s best-case limit.
This difference is attributable to the larger effective beam size 
used for our photometry/noise calculations.
On top of the confusion limit, a few sources have higher noise values
due to galactic cirrus and/or modestly worse than typical detector
performance. In particular, HD168443, located within 2.5 degrees of
the galactic plane, shows extremely high levels of cirrus noise 
(30 mJy/aperture, approximately equal to the level predicted from the
IRAS Sky Survey Atlas) and is not considered further.   

While we can directly examine the overall background noise in each of
our \70um images, the systematic error terms are more difficult to evaluate.
These errors must certainly be as large as those in the \24um data since the photospheric fitting errors should be the same for both
wavelengths. Within our data, we find that the observed dispersion of fluxes at
\70um is consistent with larger systematic errors than at \24um.
We assume that the systematic errors in the \70um data are 15\% 
of the stellar flux, about twice the dispersion in the \24um data.
Given the complex calibration required for the \70um Germanium
photoconductors, this is an excellent level of performance.

\begin{figure}
\begin{center}
\includegraphics[width=4.7in,angle=-90]{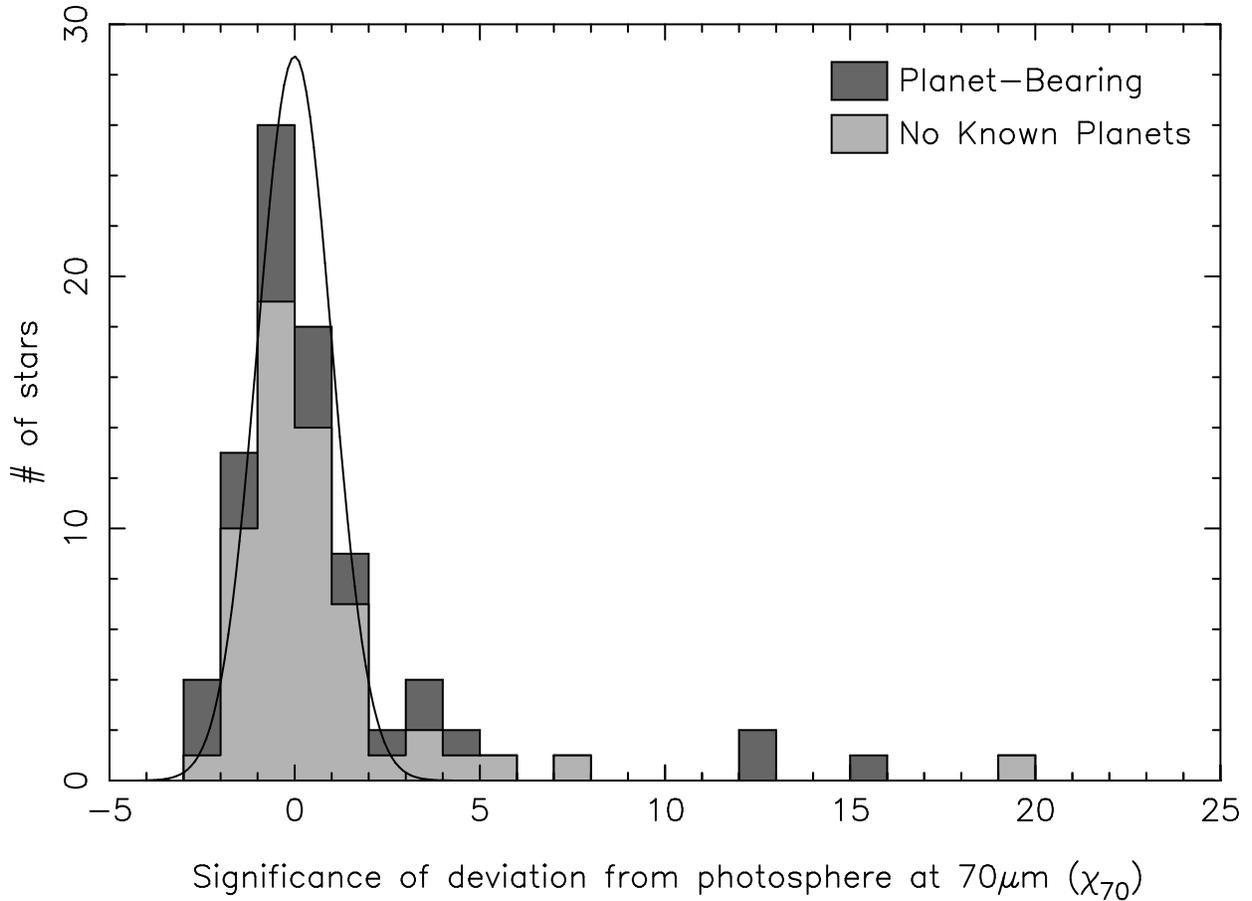} 
\end{center}
\caption{Statistical significance of deviations from the expected
stellar flux at \70um.
A gaussian distribution of errors is shown for comparison.
In order to be considered a significant \70um excess, 
$\chi_{70}$ should at least exceed the 3-$\sigma$ level.
Six of the stars with planets (dark grey) meet this criterion,
as do six of the stars without known planets (light grey).
\label{chi70}}
\end{figure}

Adding all of the noise sources (photon noise, sky background, model fitting, and residual calibration issues) together in quadrature gives us a
final noise estimate for each \70um target.
In Table \ref{planetdata} we list these noise levels, along with the
measured and the photospheric fluxes, for each of the planet-bearing stars.  
We use these noise estimates to calculate $\chi_{70}$, the statistical
significance of any IR excess
\begin{equation}\label{chi70eq}
\chi_{70} \equiv \frac{F_{70} - F_{\star}}{N_{70}}
\end{equation}
where $F_{70}$ is the measured flux, $F_{\star}$ is the expected
stellar flux, and $N_{70}$ is the noise level, all at \70um.
The resultant histogram is shown in Fig.~\ref{chi70}.
Based on this criterion, 
we find that 6 out of 25 planet-bearing stars have a 3-$\sigma$ or greater
excess at \70um: HD33636 (BD+04858), HD50554 (BD+241451), HD52265
(HR2622), HD82943, HD117176 (70 Vir), and HD128311 (GJ3860).
Of the remaining stars, 3-$\sigma$ upper limits on any excess flux
range from 0.1 to 1.9 times the stellar flux, with a median of upper
limit of 0.7 $F_{\star}$.
The focus of this paper is the planet-bearing stars; the 
non-planet stars with excess will be discussed in a separate paper on
the Volume Limited Sample \citep{bryden05}.

Fig.~\ref{images} shows the \70um images for each of the six
planet-bearing stars identified as having significant \70um excess.
The full-width half-maximum for the telescope's point spread function
at \70um is 17\arcsec; there is no evidence for any emission extended
beyond the point spread function.
The background noise fluctuations are generally smooth across the
field, with the exception of HD52265 which has obvious cirrus
contamination concentrated toward the NW corner of the image.
Within the entire $\sim$5$\times$2.5 arcmin MIPS field of view,
most of the images contain point sources other than the target star.
Some of these neighboring sources have corresponding \24um emission, 
while others are only detected at \70um.

\subsection{Notes on individual sources}

{\it HD82943:}
This is our strongest detection of an excess, with a \70um flux
15 times the stellar photosphere.  The lack of excess at \24um
requires the spectral energy distribution to rise steeply
above the photosphere between 24 and \70um. 
As the brightest 
excess detection (113 mJy), this is the most promising candidate 
for follow-up detection at longer IR and sub-mm wavelengths.
We also note that the claimed detection of $^6$Li in this star's
atmosphere could be explained by the accretion of several 
$\MEarth$
of rock \citep{Israelian01,Israelian03}.
While our observed IR excess implies that collisions could have recently
showered the star with some solid material, it is not clear whether the
magnitude of this accretion is large enough to account for 
the enhanced lithium abundance.

{\it HD33636 \& HD50554:}
These similar stars both have considerable excess at \70um
($\sim$7 times the stellar flux),
but still only emit 33 and 40 mJy respectively.
Although dim, these systems may still be detectable at sub-mm
wavelength if the dust is cold enough ($\lapp$40 K),
and even upper limits at longer wavelengths 
would greatly help to constrain the dust temperature, and hence location.

{\it HD128311:} This star shows a definite detection above the
photospheric level at \70um, but a high degree of galactic cirrus
and/or anomalous instrument artifacts make follow-up observations
important for determining the strength of the excess more accurately. 
The \spit/IRS spectrum for this source lies well below the low
quality (FQUAL=2) detection at 25 \um \ by IRAS and shows no evidence 
for an excess out to 30 \um.  
Additional photometric measurements with MIPS are currently scheduled.  

{\it HD117176 (70 Vir):}
Unlike all of our other stars with \70um excess,
the accuracy of the flux measurements for this star are
limited by systematic/fitting errors, rather than by sky noise.
Additional work to determine the source of the systematic errors
will help quantify the overall noise level for this star.

{\it HD52265:} Although this star has a high level of excess at \70um
(a factor of 4 above the stellar photosphere), this is our least
statistically significant detection of excess due to
a high level of galactic cirrus in the field.
This noise (7.4 mJy) unfortunately may not be greatly reduced with 
additional integration time.
Interestingly, this star happens to fall in the limited field of
view of the future space-based telescope COROT \citep{baglin02}.
As the only planet-bearing primary target for that mission,
its visible emission will be monitored over an extended time period for any
variability due to transiting objects. 

{\it HD142:} This star has a flux twice that of the stellar photosphere,
but the potential excess is only significant at the 2.4-$\sigma$
level.  A relatively short integration time was used (3 cycles) resulting 
in above average background noise. Additional follow-up observations of this star should  clarify whether the excess is due to dust or is just a noise 
fluctuation.

{\it 55 ($\rho$) Cnc: }
With a measured flux of $18.9 \pm 3.8$ mJy (compared to a predicted
stellar flux of 18.2 mJy), we find no evidence for an IR excess.
This is in contrast with a claimed detection of 
60 \um \ excess by ISO \citep{dominik98} at the level of $170 \pm 30$
mJy.  The original detection has
more recently been attributed to background extragalactic sources
based on sub-mm mapping of the region \citep{RayJay02}.  
We confirm the presence of two (presumably) extragalactic \70um
sources located at 08:52:37+28:20:05 and 08:52:35+28:19:18, which
correspond to sub-mm sources 1 and 3 from \citet{RayJay02}.
However, the \70um flux from these neighboring sources is small; a
beam large enough to enclose them makes little appreciable change
in the total integrated flux.

{\it $\rho$ CrB \& HD210277:} Along with 55 Cnc, these stars were identified 
as having extended emission via coronographic measurements at visible
wavelengths \citep{trilling00}. We find no evidence for excess emission around either star at 24 or \70um.

\section{Correlation of Excess with System Parameters}\label{parameters}

Here we examine whether IR excess is correlated 
with stellar properties (age, metallicity, spectral type) 
and planet orbital characteristics (planet mass, 
semi-major axis, eccentricity).
The planet and star properties are summarized in Tables 
\ref{planetparam} and \ref{starparam}.

The planetary systems considered here all have at least one planet
with mass greater than Jupiter's and with semi-major axis smaller
than Jupiter's. 
Considering this sample in aggregate, there is a clear trend 
between the presence of this type of planet and the detection of IR
excess. Within our sample we have found 6 stars with excesses out of
the 25 planet-bearing stars for a detection rate of $24\pm10$\%. 
This is much less than the excess detection rates for young stars;
\citet{Habing01} observe that 60\% of A-K stars younger than 400 Myr 
have excess 60 \micron\ emission, while \citet{Rieke05} similarly 
find that half of young A stars have excess \24um emission.  
Analysis of IRAS and ISO data suggests
that the fraction of stars with excesses drops considerably with
increasing age, however, particularly beyond 1 Gyr \citep{Habing01,
Decin03}.   
The nominal IRAS statistics suggest that excesses are found around
just 10-15\% of all main-sequence stars \citep{Backman93},
a factor of two less than our detection rate.
If ages older than 1 Gyr are considered, the disparity in rates
becomes more significant. 
ISO surveys found excesses around only
$10\pm$3\% of stars older than 400 Myr \citep{Decin00,Habing01}. 
This detection rate is similar to that found within our sample of
stars not known to have planets \citep{bryden05}.  
Within this group, $10\pm4$\% (6 out of 58) of the stars exhibit
excess IR emission at \70um. 
Among stars older than 1 Gyr, the detection rate for stars with
planets is more than double that for stars without.
This tentative result would be consistent with the
suggestion of \cite{Dominik03} that excesses due to Kuiper Belt
material can have very long lifetimes. 
The apparent increase in frequency of dust disks around planet-bearing
stars over stars without known planets implies that searches for dust
disks might benefit from targeted searches of planet-bearing stars.
The converse would also be true - 
planet searches, vis radial velocity or otherwise, may benefit
from specifically targeting stars with IR excesses.
This correlation is preliminary and
will be examined closely as additional stars in this survey are observed.  

\subsection{Planet Orbital Characteristics}

\begin{figure}
\begin{center}
\includegraphics[width=4.7in,angle=-90]{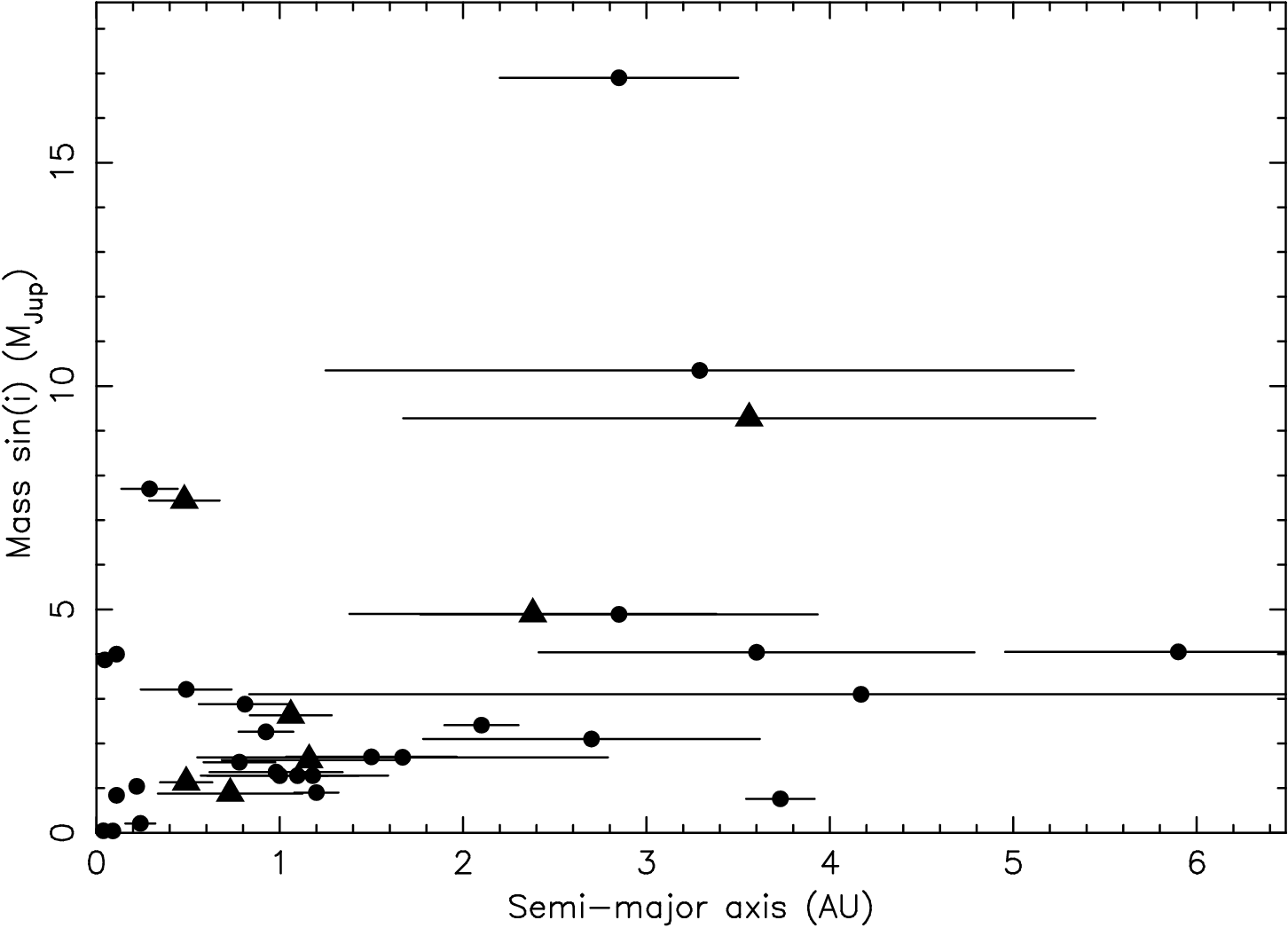} 
\end{center}
\caption{Orbital characteristics for the survey stars.
Minimum planet mass ($M_p \rm{sin}i$) is plotted versus each planet's 
semi-major axis ($a_p$).  The lines extend from periapse 
($a_p ( 1 - e_p)$) to apoapse ($a_p (1 + e_p)$).
The planets which are associated with \70um excess are marked
as triangles, while those without are circles.
\label{planetfig}}
\end{figure}

What kind of planets do stars with debris disks possess? Table
\ref{planetparam} and Fig.~\ref{planetfig} 
summarize the properties of the planetary systems. 
Two of the systems with strong excesses have large planets on
distant, eccentric orbits (9.3 $\MJup$ at 3.6 AU for HD33636 and 4.9
$\MJup$ at 2.4 AU for HD50554).  The system with strongest excess,
HD82943, has two moderate-sized planets with masses $\sim$1 $\MJup$ at
0.73 and 1.16 AU.
Overall, however, the distribution of planets associated
with IR excess is little different than for those without an
excess. For every type of planet with IR excess, there are
counter-examples without excess: for example, HD39091 has a large planet on a 
distant, eccentric orbit yet shows no dust emission.  
The only possible trend is that no systems with short-period planets
($<$0.25 AU) exhibit any IR excess, though the small-number
statistics limit the strength of this conclusion.

Planets should affect the diffuse material in their immediate
vicinity. The region under the direct gravitational
influence of a planet at $a_p$ with eccentricity $e_p$
has a half-width of $\sqrt{4e_p^2 a_p^2/3 + 12 r_{H}^2}$,
where $r_{H} \equiv a_p (\Mp/3\Mstar)^{1/3}$ is the Hill radius 
\citep[e.g.][]{brydennept}.
For the planet orbiting HD33636 
($M_p {\rm sin} i$=9.28$\MJup$, $a_p$=3.56 AU, $e_p$=0.53)
this directly perturbed region extends 
from 1 to 6 AU where we have ignored the unknown inclination effect
which enters only weakly into the Hill radius. This zone would be
cleared of any large objects that might provide a local source of dust
generation.  Although dust could drift toward the planet from a distant
belt of source bodies, Jupiter-mass planets are efficient at
gravitationally scattering this dust from their neighborhood
\citep{Moro05}.
Outside of the planet's region of direct influence, 
substantial Kuiper Belt material
could be present at distances $\gapp$10 AU and effective temperatures
below 100 K, consistent 
with the cool excess (no \24um emission) seen toward these
stars. Additional \spit\ and ground-based submillimeter data are
required to constrain the temperature and the physical distribution
of this material \citep{Sheret04}. 

\subsection{Stellar Properties}\label{agesec}

\begin{figure}
\begin{center}
\includegraphics[width=4.7in,angle=-90]{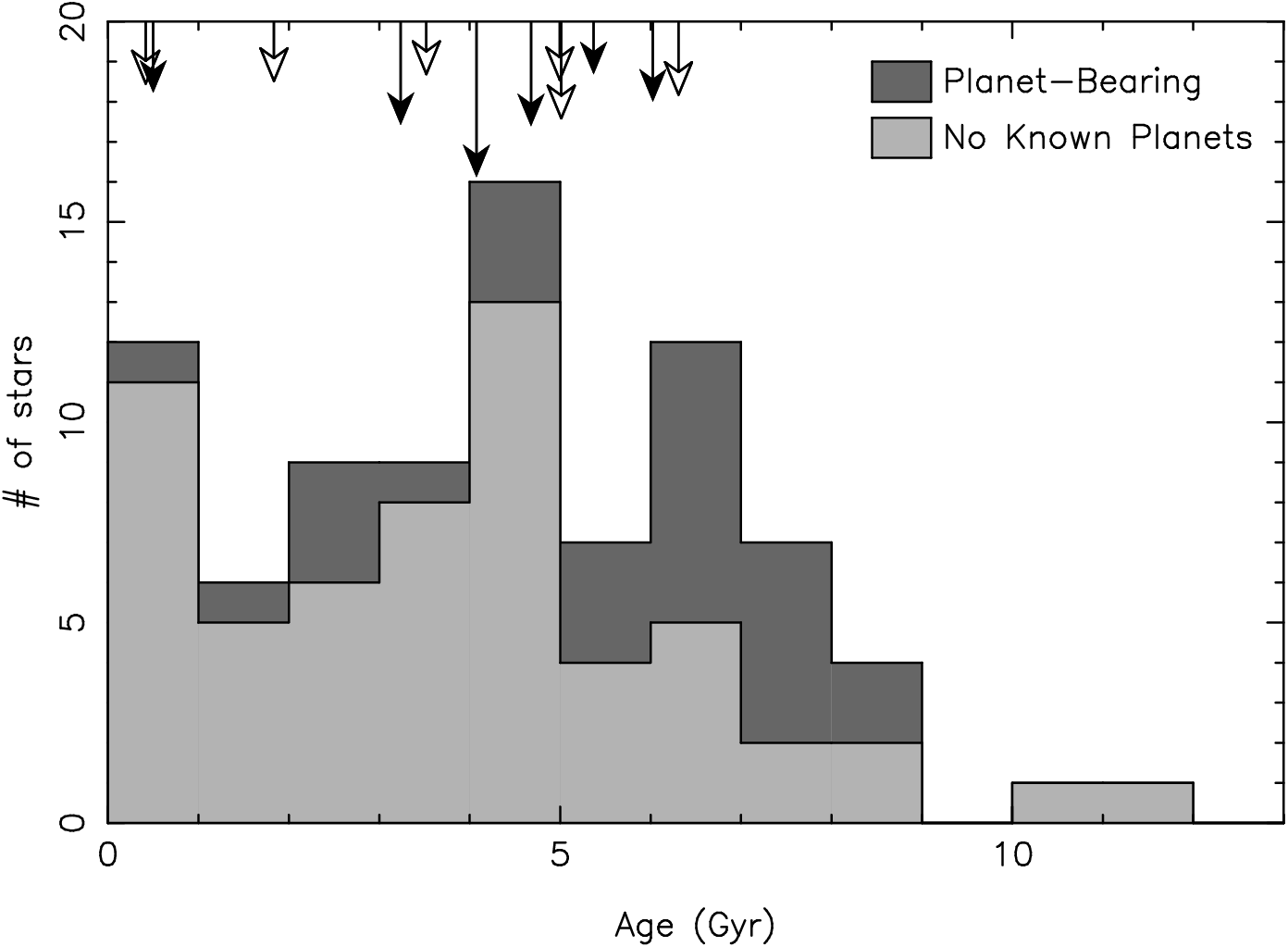} 
\end{center}
\caption{Distribution of stellar ages.
Stars with planets are shaded as dark grey
while stars without known planets are designated with light grey.
The ages of stars with \70um excess are flagged as arrows at the top
of the plot.  For planet-bearing stars the arrows are filled; for stars
without known planets they are open. The length of the arrow is an
indicator of the strength of \70um excess.
\label{ages}}
\end{figure}

Stellar ages cited in the literature for these stars
span at least a factor of two, highlighting the difficulty in 
determining ages for mature, main sequence stars.  
Whenever possible, we adopt ages
from the compilation of \cite{Wright04} which provides a uniform
tabulation for 1200 stars based on Ca II H\&K line strengths.
While Table \ref{starparam} lists the ages for the planet-bearing stars,
Fig.~\ref{ages} gives a histogram for the larger sample, both with and
without planets.

There is no obvious age difference between the overall planet-bearing 
and non-planet bearing samples; their median ages are 6 and 4 Gyr respectively.
The one outlier among the planet-bearing sample is HD128311, which
has no well-determined age. It is listed as a possible member of the
UMa moving group \citep{King03} based on its chromospheric activity
and its X-ray detection in the ROSAT All Sky Survey. Both of these
indicators suggest an age of 0.5-1 Gyr. In Fig.~\ref{ages}, the stars
with excess are flagged with arrows, solid for planet-bearing stars
and open for those without.   
With the exception of HD128311, the planet-bearing stars with excesses
are all much older than a billion years. The planet-bearing excess
stars are slightly younger than the planet-bearing sample as a whole
(4.4 vs. 6.0 Gyr), but the difference is not statistically
significant. Any inverse correlation of age with amount of dust is
much weaker than that observed in stars younger than 1 Gyr
\citep[e.g.][]{Decin03, Rieke05}.

\begin{figure}
\begin{center}
\includegraphics[width=4.7in,angle=-90]{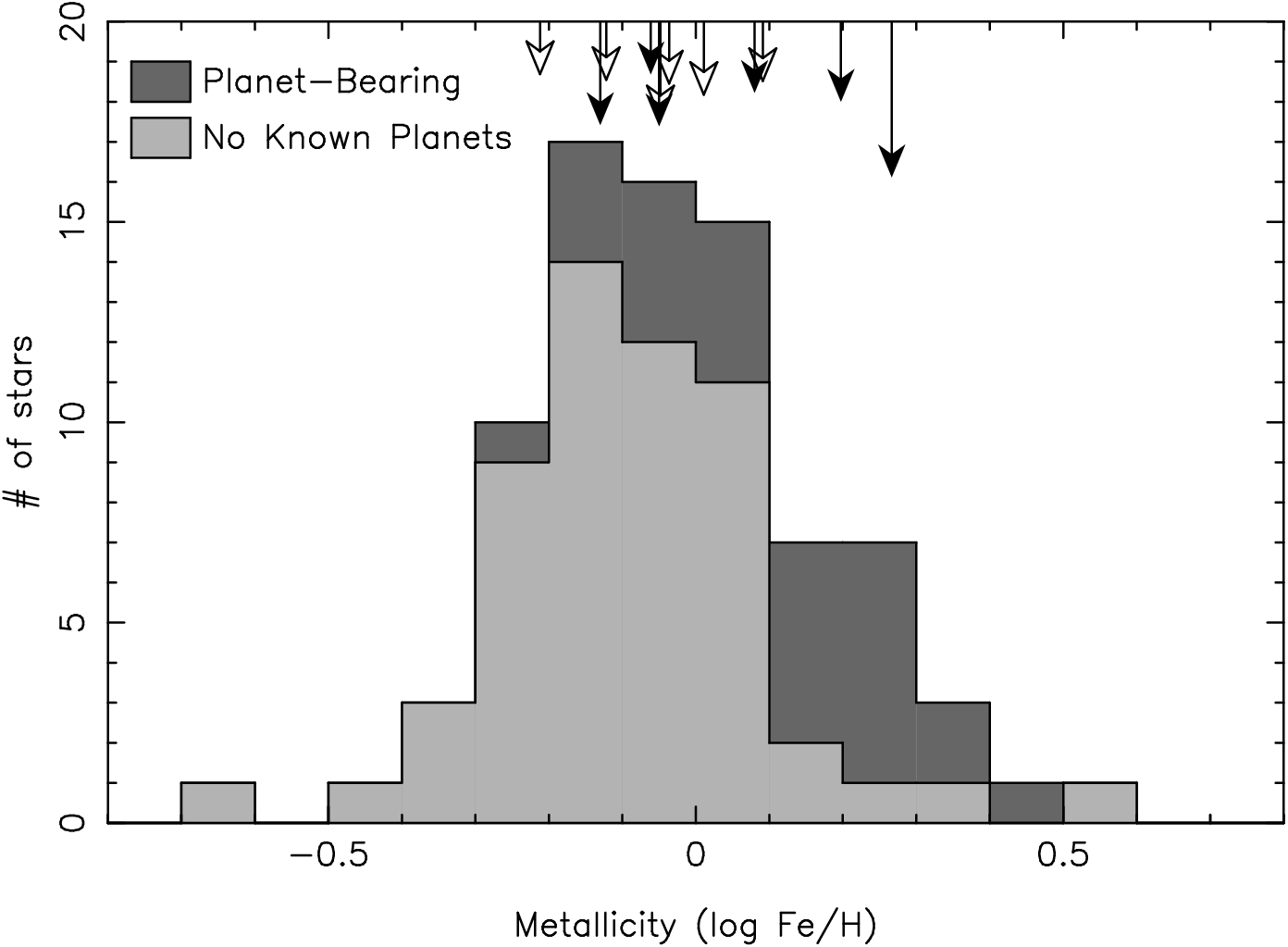} 
\end{center}
\caption{Distribution of stellar metallicities.
Stars with planets are shaded as dark grey
while stars without known planets are designated with light grey.
The metallicities of stars with \70um excess are flagged as arrows at
the top of the plot. For planet-bearing stars the arrows are filled;
for stars without known planets they are open. The length of the
arrow is an indicator of the strength of \70um excess.
\label{metals}}
\end{figure}

Fig.~\ref{metals} presents a similar histogram for the metallicity 
of stars with/without planets, again marking IR excess stars with
arrows. The distribution of metallicities for planet-bearing stars (dark grey)
is clearly different than for those without planets (light grey).
The typical metallicity of stars with planets is well known to be greater than
solar \citep{Wright04, Fischer03} so the fact that the average
metallicity of these stars is slightly greater than solar,
[Fe/H]$=0.13 \pm 0.05$ with a dispersion of 0.19, is no surprise.  
Although the planet-bearing stars with excess have an average 
metallicity above solar, their average [Fe/H]=0.05 is below
the planet-bearing average. Comparing the stars with and without
planets to their own samples, there is no evidence for higher 
metallicity resulting in a greater amount of IR emitting dust.

\section{Analysis of Detected Dust}\label{ldsec}

Unfortunately, our detections of IR excess provide only 
limited information about the properties of the dust in each system.
Ideally, we would like to determine the dust's temperature,
mass, grain sizes, and overall distribution.
Because none of our sources are resolved (see Fig.~\ref{images}), 
we have no information  on whether the dust disk has any asymmetric structure such as a warp, a centering offset, or general clumpiness.
Still, we can use the spectral energy distribution (SED) to provide 
some constraints on the dust properties.
The most direct constraint provided by the SED is on the dust 
temperature, which then can be converted into a radial distance
from the central star. 
As seen in Fig.~\ref{sedboth}, \24um measurements provide an upper
limit on the dust temperature, while sub-mm limits set a lower limit.
Because we usually have no information longward of \70um, however,
only an upper limit on the temperature (or an inner 
limit on the dust's orbital location) can be derived.
If a dust temperature is assumed, the observed flux can
be translated into the total dust disk luminosity relative 
to its parent star. 

For disks with detections of IR excess, a minimum dust
luminosity can be calculated.
Using the blackbody formula for flux ($F \propto L / T^4 (e^{h\nu/kT}-1)$),
for long wavelength radiation (i.e.\ on the Rayleigh-Jeans tail of 
the stellar blackbody curve), the ratio of dust to stellar fluxes becomes
\begin{equation}
\frac{F_{{\rm dust}}}{F_{\star}} = 
\frac{L_{\rm dust}}{L_{\star}} \;
\frac{h\nu T_{\star}^3}{kT_{\rm dust}^4 (e^{h\nu/kT_{\rm dust}} -1)}
\end{equation}
The minimum disk luminosity as a function of \70um dust flux can be
obtained unambiguously by setting the emission peak at \70um
(or, equivalently, $T_{\rm dust} = 52.5$K) : 
\begin{equation}\label{ldeq}
\frac{L_{\rm dust}}{L_{\star}}({\rm minimum}) = 10^{-5} 
\; \left(\frac{5600 \; {\rm K}}{T_{\star}}\right)^3
\; \frac{F_{70, {\rm dust}}}{F_{70, \star}}
\end{equation}

\begin{figure}
\begin{center}
\includegraphics[width=4.7in,angle=-90]{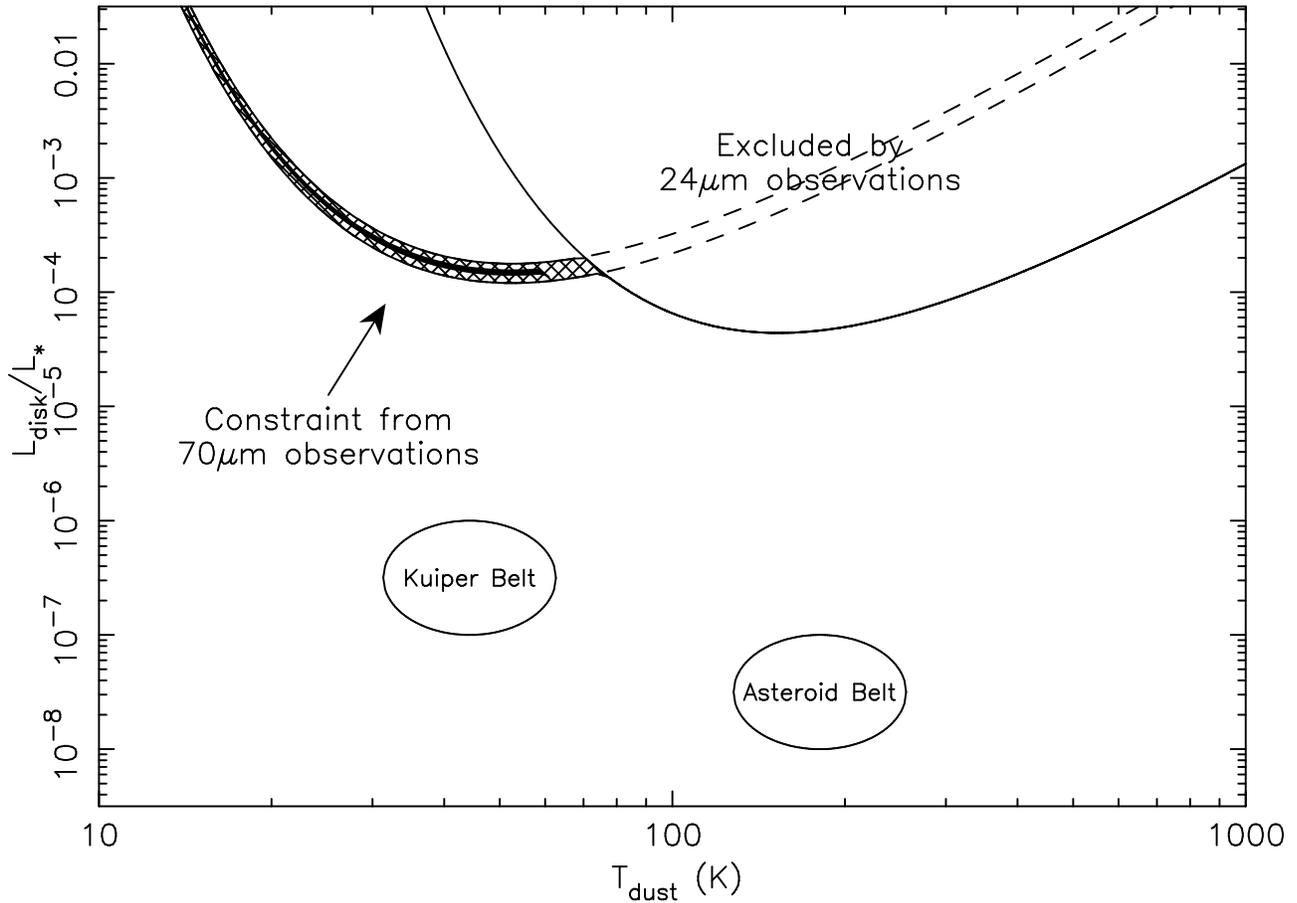} 
\end{center}
\caption{Constraints on the temperature and total luminosity of 
the dust around HD82943, as provided by 24 and \70um MIPS observations.
The area to the upper right (hot temperature, bright luminosity) is ruled 
out by the \24um upper limit on the dust emission.
The 3-$\sigma$ limits around \70um detection confine the dust 
characteristics within the cross-hatched region.
The solid-filled region is defined by 1-$\sigma$ error limits.
Without any longer wavelength information, there is no lower limit on the 
dust temperature or upper limit on the dust luminosity.
The approximate characteristics of the asteroid and Kuiper
belts are shown for comparison.
\label{ldlstar}}
\end{figure}

Consistent with this formula, Fig.~\ref{ldlstar} shows constraints on 
$\ld$ and $T_{\rm dust}$ for HD82943, the planet-bearing star with \70um
flux a factor of 15 above the stellar photosphere.
The lines in this figure are based on the 3-$\sigma$ limits to the
observed 24 and \70um fluxes.
$T_{\rm dust}$ is meant to signify the typical emitting temperature for the
dust; in reality some range of temperatures will be found in any given
system.
For this star, our strongest detected excess, the minimum $\ld$ is
relatively high - $1.4\times 10^{-4}$ - more than two orders of
magnitude above estimates for the Kuiper belt's luminosity.
(A listing of minimum $\ld$ for the other planet-bearing stars
is given in Table \ref{planetdata}.)
The dust temperature is well-constrained by the lack of \24um excess
to be $\lapp$70 K at the 3-$\sigma$ level (the cross-hatched region in
Fig.~\ref{ldlstar}) or $\lapp$60 K at the 1-$\sigma$ level (the
solid-filled region). Dust at lower temperatures showing greater emission at longer wavelengths cannot be ruled out.

\begin{figure}
\begin{center}
\includegraphics[width=4.7in,angle=-90]{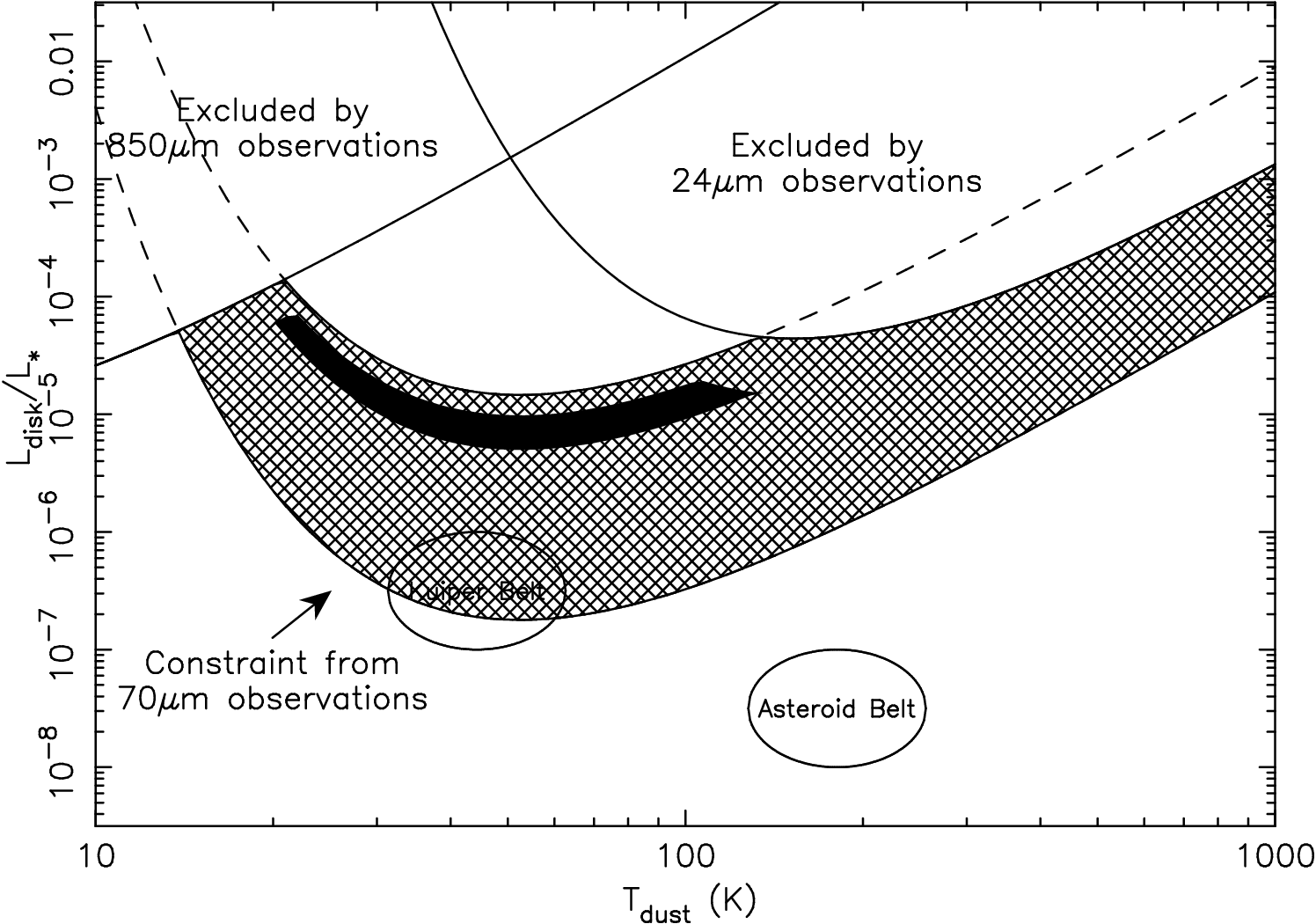} 
\end{center}
\caption{Constraints on the temperature and total luminosity of 
the dust around HD117176, as provided by 24 and \70um MIPS data
combined with sub-mm observations.
The upper right is ruled out by the \24um upper limit,
while the upper left is ruled out by the sub-mm 
\citep{Greaves04}.
Based on the \70um data, the dust temperature and luminosity are
confined to the shaded region. 
The 3-$\sigma$ limits for the three wavelengths confine the dust 
characteristics within the cross-hatched region, while 
the solid-filled region is defined by 1-$\sigma$ error limits.
Again, the approximate characteristics of the asteroid and Kuiper
belts are shown for comparison.
Note that although the formal 3-$\sigma$ error limits extend as low as
the Kuiper belt's luminosity, the Kuiper belt itself would be too
faint to detect. 
\label{ldlstar2}}
\end{figure}

Sub-mm observations are critical to constraining
the dust properties and thus the overall mass of the material
responsible the the excess in these stars. 
Fig.~\ref{ldlstar2} shows the limits on the dust characteristics
for a case where sub-mm data is available to help constrain the
dust temperature on the low end.
(\citet{Greaves04} observed HD117176 at 850 \um, measuring 
a flux of $2.3 \pm 2.4$ mJy.)
Note that in order to set the most strict limits, we assume that the
grains are small enough that their emissivity drops off as 
$\lambda^{-2}$ for long-wavelength radiation \citep[see
e.g.][]{Draine84,Wyatt02}. 
The sub-mm measurement does set a lower limit for the dust
temperature, but
because the \70um emission is not as strong as for the previous star,
the upper limit on temperature is less strict.
Overall, the possible range of dust temperature and luminosity
are much better constrained: 
1-$\sigma$ limits (the solid-filled region in Fig.~\ref{ldlstar2}) give 
$T_{\rm dust} \approx 20$ to 100 K, $\ld \approx 10^{-5.2}$ to $10^{-4.2}$.
For a temperature of 50 K, this disk brightness corresponds to a total
dust surface area of $\sim$$10^{25}$-$10^{26}$ cm$^2$ or, equivalently,
a dust mass of $\sim$$10^{-6}$-$10^{-5} \MEarth$ in 100 \micron \ grains.
The mass contained in larger planetesimals is even less certain.

For stars with no detected emission, a 3-$\sigma$ upper limit on the
fluxes leads to upper limits for $\ld$ of $10^{-6}$
to $10^{-5}$, assuming a dust temperature of $\sim$50 K
(again see Table \ref{planetdata}).
For these stars we are setting limits of
$\sim$10-100 times the level of dust in our solar system. Deeper
integrations on these stars may result in detections of fainter
excesses or improved upper limits, but extragalactic confusion or
cirrus will set limits on the amount of detectable excess for these
relatively distant stars. Non-planet bearing stars in the VLS are
brighter and typically located in more cirrus-free regions, so that
our $\ld$ limits for these objects tend to be more
sensitive than for the planet-bearing stars reported here. 

\section{Conclusions}\label{conclusions}

\spit/MIPS observations of 26 main-sequence FGK stars which are
known to have planetary companions reveal \70um dust emission from 6
debris disks.  These are the first examples outside the solar system of stars which have both well-confirmed planets and well-confirmed IR excesses.
We conclude from this study that excesses at \70um are
at least as common around planet-bearing F5-K5 stars as they are
around the A-F stars first detected by IRAS. 

Our finding is in contrast with \citet{Greaves04}, who 
found no overlap between stars with disks and stars with planets. 
In order to look for dust/planet systems, they observed a sample of 8 planet-bearing stars in the sub-mm, with no detections.
The disks identified by \spit, however, are considerably less
massive than those detectable in the submillimeter. 
For example, in the case of $\epsilon$ Eri, the ratio of 60 
\um \ IRAS emission (1,600 mJy) to 850 \um \ emission (40 mJy) is 40:1. 
If this same ratio applies to the sources seen here which have a typical 
\70um brightness of 35 mJy, the predicted submillimeter 
flux density would be $\sim$1 mJy which is just the 1-$\sigma$ 
noise limit for the SCUBA survey. Six of the eight sub-mm sources are also in our target sample, allowing a direct comparison of \spit\ with SCUBA.
In contrast with the sub-mm non-detections, five photospheres were
observed with SNR$\gapp$5, and one star (HD117176, 70 Vir) was 
found to have significant \70um excess. While SCUBA follow-up observations 
of the sources detected here will be very important for constraining the 
temperature and total mass of dust, conclusions about the incidence of 
disks around planet-bearing stars require \spit's higher sensitivity.

While excesses at \70um are relatively frequent in our survey, we find
no evidence for excess at \24um around any of the planet-bearing stars.
The lack of \24um emission in these systems is
consistent with the previously known lack of warm material toward
stars older than a few 100 Myr \citep{Aumann91, Fajardo00, Laureijs02}. 
Assuming peak \24um emission from 150 K dust and the observed dispersion
($F_d/F_{\star}<0.07$), we can set a 3-$\sigma$ upper limit on $\ld$
for hotter dust of $5 \times 10^{-5}$ which is well above the
$10^{-7}$ level of emission from our asteroid belt,  but well below
the level seen toward young A stars such as $\beta$ Pic or Fomalhaut
($\alpha$ PsA).  
As described above, the lack of warm material is consistent with
its destruction or dispersal by the gravitational effect of planets in
the 1-5 AU region \citep{Liou99}. 

For stars with a single measurement of excess at \70um, 
the dust properties are not well constrained, but are generally 
consistent with Kuiper belt configurations -
distances of several tens of AU and  temperatures $\sim$50 K. The
observed dust luminosities (relative to the central stars'), however,
are much brighter, exceeding the Kuiper belt's  $\ld$ by factors of
$\sim$100.  

Within the planet-bearing sample there is little to no correlation
of \70um excess with stellar age, metallicity, or spectral type. But 
there is a strong hint of a positive correlation between both the
frequency and the magnitude of dust emission with the presence of
known planets. As more \spit\ observations of planet-bearing and
non-planet-bearing stars are made, as part of this survey and in other
programs, these statistics will become much better defined. Additional
radial velocity studies of stars with excesses but without presently
known planets will test the predictive power of this correlation. 

\acknowledgments {
We would like to thank Debra Fischer and Geoff Marcy for
discussions on stellar metallicity, 
updates on RV planet searches, and
commentary on $\epsilon$ Eri's variability.
This publication makes use of data products from the Two Micron All
Sky Survey (2MASS), as well as from IPAC/ibis, SIMBAD, and VIZIER;
Samantha Lawler assisted in the compilation of data from these sources.
We acknowledge technical support from the Center for Long-Wavelength Astrophysics at JPL, particularly from J. Arballo and T. Thompson.

The {\it Spitzer Space Telescope} is operated by the Jet Propulsion
Laboratory, California Institute of Technology, under NASA contract 1407. 
Development of MIPS was funded by NASA through the Jet Propulsion
Laboratory, subcontract 960785.  Some of the research described in
this publication was carried out at the Jet Propulsion Laboratory,
California Institute of Technology, under a contract with the National
Aeronautics and Space Administration.  

Finally, we note that much of the preparation for the observations described here was carried out by Elizabeth Holmes, who passed away 
in March 2004.  This work is dedicated to her memory.}

\newpage
\begin{table}
\vskip -0.2in
\begin{center}
\begin{tabular}{||cc|ccc||} 
\hline \hline
 HD \#  & name  & $M_p {\rm sin}i \ (M_{\rm Jup})$ 
   & $a_p$ (AU) & $e_p$  \\
\hline 
142 & GJ4.2A  & 1.36 & 0.98 & 0.37 \\
1237 &  GJ 3021  & 3.21 & 0.49 & 0.505 \\
13445 & Gl 86  & 4 & 0.11 & 0.046  \\
17051 & $\iota$ Hor, HR 810 & 2.26 & 0.925 & 0.161  \\
27442 & GJ167.3 & 1.28 & 1.18 & 0.07  \\
33636$^*$ &  BD+04858  & 9.28 & 3.56 & 0.53  \\
39091 &  GJ 9189  & 10.35 & 3.29 & 0.62  \\
50554$^*$ &  BD+241451  & 4.9 & 2.38 & 0.42  \\
52265$^*$ &  HR 2622  & 1.13 & 0.49 & 0.29  \\
75732 &  55 ($\rho$) Cnc  & 0.045 & 0.038 & 0.174  \\
    &  & 0.84 & 0.11 & 0.02  \\
    &  & 0.21 & 0.24 & 0.34      \\
    &  & 4.05 & 5.9 & 0.16      \\
82943$^*$ & BD-112670 & 0.88 & 0.73 & 0.54 \\
 &  & 1.63 & 1.16 & 0.41      \\ 
95128 & 47 Uma  & 2.41 & 2.1 & 0.096  \\
 &  & 0.76 & 3.73 &  $<$ 0.1       \\
114783 & GJ 3769 & 0.9 & 1.2 & 0.1  \\
117176$^*$ & 70 Vir & 7.44 & 0.48 & 0.4  \\
120136 & $\tau$ Boo, GJ 527A & 3.87 & 0.0462 & 0.018   \\
128311$^*$ &  GJ 3860  & 2.63 & 1.06 & 0.21 \\
134987 &  23 Lib  & 1.58 & 0.78 & 0.25  \\
143761 &  $\rho$ CrB  & 1.04 & 0.22 & 0.04  \\
145675 &  14 Her  & 4.89 & 2.85 & 0.38  \\
160691 &  $\mu$ Arae, GJ 691 & 0.042 & 0.09 & 0  \\
 &  & 1.7 & 1.5 & 0.31  \\
 &  & 3.1 & 4.17 & 0.8     \\
168443 &  GJ 4052  & 7.7 & 0.29 & 0.529  \\
 &  & 16.9 & 2.85 & 0.228  \\   
169830 &  HR 6907  & 2.88 & 0.81 & 0.31 \\
 &  & 4.04 & 3.6 & 0.33      \\
177830 &  GJ 743.2  & 1.28 & 1 & 0.43  \\
186427 & 16 Cyg B  & 1.69 & 1.67 & 0.67  \\
210277 & GJ 848.4  & 1.28 & 1.097 & 0.45 \\
216437 & $\rho$ Ind & 2.1 & 2.7 & \hspace{0.2in} 0.34 \hspace{0.2in}  \\
\hline \hline
\end{tabular} 
\end{center}
$^*$star with excess \70um emission
\caption{Orbital characteristics of the known planets surrounding
stars observed by Spitzer.  Also see Fig~\ref{planetfig} for a
graphical display of some of this information.} 
\label{planetparam}
\end{table}

\newpage
\begin{table}
\hskip -0.6in
\begin{tabular}{||c|ccc|ccccccc||} 
\hline \hline
  & & \24um  & & & & & \70um &  & & \\
 HD \#  &  $F_{\rm MIPS}$  &  $F_{\star}$  &  $F_{\rm MIPS}/F_{\star}$  
&  $F_{\rm MIPS}$  &  $F_{\star}$  &  $F_{\rm MIPS}/F_{\star}$ & SNR 
&  $\chi_{70}$ & $F_{\rm dust}$$^{\dagger}$ & $\ld$ \\ 
\hline 
142 & 117.3 & 123.3 & 0.95 & 26.4 $\pm$ 5.1 & 13.9 & 1.9 & 5.6 & 2.4 & & $<$2.0$\times 10^{-5}$ \\
1237 & 82.9 & 88.7 & 0.94 & 10.0 $\pm$ 2.9 & 10.1 & 1.0 & 4.0 & 0.0 & & $<$9.0$\times 10^{-6}$ \\
13445 & 162.9 & 166.9 & 0.98 & 3.9 $\pm$ 6.3 & 19.0 & 0.2 & 0.7 & -2.4 & & $<$4.1$\times 10^{-6}$ \\
17051 & 166.8 & 161.7 & 1.03 & 20.1 $\pm$ 4.1 & 18.1 & 1.1 & 6.4 & 0.5 & & $<$6.7$\times 10^{-6}$ \\
27442 & 1147.7 & 1222.7 & 0.94 & 126.5 $\pm$ 21.9 & 139 & 0.9 & 19.2 & -0.6 & & $<$7.8$\times 10^{-6}$ \\
33636$^*$ & 42.9 & 44.1 & 0.97 & 33.1 $\pm$ 2.3 & 5.0 & 6.7 & 15.4 & 12.4 &
32.4 & { 4.9$\times 10^{-5}$} \\
39091 & 139.9 & 150.4 & 0.93 & 21.5 $\pm$ 3.6 & 17.0 & 1.3 & 8.3 & 1.3 & & $<$9.0$\times 10^{-6}$ \\
50554$^*$ & 47.1 & 51.2 & 0.92 & 39.5 $\pm$ 2.8 & 5.8 & 6.8 & 14.9 &
  12.1 & 38.8 & 4.4$\times 10^{-5}$ \\ 
52265$^*$ & 74.3 & 70.7 & 1.05 & 31.7 $\pm$ 7.4 & 8.0 & 4.0 & 4.3 &
  3.2 & 27.3& 2.9$\times 10^{-5}$ \\ 
75732 & 172.8 & 162.7 & 1.06 & 18.9 $\pm$ 4.5 & 18.2 & 1.0 & 5.3 & 0.2 & & $<$8.3$\times 10^{-6}$ \\
82943$^*$ & 66.5 & 66.5 & 1.00 & 113.3 $\pm$ 6.8 & 7.5 & 15.1 & 17.0 &
  15.7 &  122 & 1.2$\times 10^{-4}$ \\ 
95128 & 259.5 & 265.9 & 0.98 & 29.1 $\pm$ 6.0 & 30 & 1.0 & 7.3 & -0.1 & & $<$4.8$\times 10^{-6}$ \\
114783 & 45.2 & 52.1 & 0.87 & 6.2 $\pm$ 2.2 & 6.0 & 1.0 & 3.2 & 0.1 & & $<$1.4$\times 10^{-5}$ \\
117176$^*$ & 373.6 & 395 & 0.95 & 77.4 $\pm$ 10.2 & 44.8 & 1.7 & 10.2 &
  3.2 & 37.6 & 1.0$\times 10^{-5}$ \\ 
120136 & 338.2 & 321.7 & 1.05 & 30.9 $\pm$ 7.9 & 36.3 & 0.9 & 5.4 & -0.7 & & $<$3.7$\times 10^{-6}$ \\
128311$^*$ &  67$^\ddagger$  & 73.5 & 0.91 & 26.5 $\pm$ 3.9 & 8.4 & 3.2
  & 7.1 & 4.6 & 20.8 &  3.0$\times 10^{-5}$ \\ 
134987 & 78.8 & 68.3 & 1.15 & 3.1 $\pm$ 4.5 & 7.7 & 0.4 & 0.7 & -1.0 & & $<$1.1$\times 10^{-5}$ \\
143761 & 201.8 & 192.5 & 1.05 & 27.8 $\pm$ 6.1 & 21.7 & 1.3 & 5.4 & 1.0  & & $<$9.6$\times 10^{-6}$ \\
145675 & 92.4 & 106.4 & 0.87 & 9.2 $\pm$ 2.5 & 12 & 0.8 & 5.4 & -1.1 & & $<$4.8$\times 10^{-6}$ \\
160691 & 270.6 & 273.5 & 0.99 & 24.1 $\pm$ 8.5 & 30.8 & 0.8 & 3.4 & -0.8 & & $<$6.1$\times 10^{-6}$ \\
168443 & 55.6 & 53.3 & 1.04 & 9.6 $\pm$ 29.6$^\diamond$  & 6.0 & 1.6 & 0.3 & 0.1 & & $<$1.5$\times 10^{-4}$ \\
169830 & 99.0 & 104.8 & 0.94 & 4.7 $\pm$ 5.9 & 11.9 & 0.4 & 0.8 & -1.2 & & $<$6.5$\times 10^{-6}$ \\
177830 & 88.6 & 104.3 & 0.85 & 4.4 $\pm$ 3.5 & 11.8 & 0.4 & 1.5 & -2.1 & & $<$3.2$\times 10^{-6}$ \\
186427 & 93.0 & 103.1 & 0.90 & 3.0 $\pm$ 5.6 & 11.6 & 0.3 & 0.5 & -1.5 & & $<$4.3$\times 10^{-6}$ \\
210277 & 83.5 & 91.9 & 0.91 & 8.0 $\pm$ 2.9 & 10.4 & 0.8 & 3.2 & -0.8 & & $<$5.1$\times 10^{-6}$ \\
216437 & 107.5 & 102.1 & 1.05 & 9.5 $\pm$ 3.9 & 11.5 & 0.8 & 2.7 & -0.5 & & $<$8.4$\times 10^{-6}$ \\
\hline \hline
\end{tabular} 
$^*$star with excess \70um emission

$^\dagger$dust fluxes have been color corrected by 15\%, appropriate for
$\sim$50K emission 

$^\ddagger$\24um flux is from IRAS 25 \um 

$^\diamond$high galactic cirrus contamination

\caption{Measured and predicted fluxes at 24 and \70um for the
planet-bearing stars, in mJy. 
For the \70um data, we also list the signal-to-noise ratio (SNR),
the significance level of any excess ($\chi_{70}$; see
Eq~\ref{chi70eq}),
and $\ld$, the minimum disk luminosity (see Eq~\ref{ldeq} in \S\ref{ldsec}).
}
\label{planetdata}
\end{table}

\newpage
\begin{table}
\begin{center}
\begin{tabular}{||c|cccc||} 
\hline \hline
 HD \# & Spectral Type & [Fe/H] & dist. (pc) & age (Gyr) \\
\hline 
142 & G1 IV & 0.02 & 25.64 & 2.9 \\
1237 & G6 V & 0.10 & 17.62 & 2.8 \\
13445 & K1 V & -0.19 & 10.95 & 5.6 \\
17051 & G0 V & 0.09 & 17.24 & 2.4 \\
27442 & K2 IVa & 0.21 & 18.23 & 6.6 \\
33636$^*$  & G0 & -0.13 & 28.69 & 3.2 \\
39091 & G1 V & 0.11 & 18.21 & 5.6 \\
50554$^*$  & F8  & -0.05 & 31.03 & 4.7 \\
52265$^*$  & G0 & 0.20 & 28.07 & 6.0 \\
75732 & G8 V & 0.31 & 12.53 & 6.5 \\
82943$^*$  & G0 & 0.27 & 27.46 & 4.1 \\
95128 & G0 V & -0.02 & 13.91 & 6.0 \\
114783 & K0 V & -0.13 & 20.43 & 4.4 \\
117176$^*$  & G2.5Va & -0.06 & 18.11 & 5.4 \\
120136 & F7 V & 0.24 & 13.51 & 1.9 \\
128311$^*$ & K0 & 0.08 & 16.57 & 
0.5$^\dagger$ \\
134987 & G5 V & 0.28 & 25.65 & 7.8 \\
143761 & G0 V & -0.23 & 17.43 & 7.4 \\
145675 & K0 V & 0.41 & 18.15 & 6.9 \\
160691 & G3 IV-V & 0.29 & 15.28 & 6.7 \\
168443 & G5 V & -0.01 & 37.88 & 8.5 \\
169830 & F8 & 0.14 & 36.32 & 7.2 \\
177830 & K0 & 0.36 & 59.03 & 8.5 \\
186427 & G3 V & 0.06 & 21.41 & 7.4 \\
210277 & G0V & 0.23 & 21.29 & 6.8 \\
216437 & G2.5IV & 0.20 & 26.52 & 7.2 \\
\hline \hline
\end{tabular} 
\end{center}
$^*$star with excess \70um emission

$^\dagger$see discussion of HD128311's age in \S\ref{agesec}
\caption{Spectral type, metallicity, distance, and age
for planet-bearing stars observed by Spitzer.}
\label{starparam}
\end{table}

\begin{figure}
\begin{center}
\end{center}
\caption{\70um images for the six planet-bearing stars with
excess emission.
The contrast scale is the same for each image.
The images are $\sim$2 arcmin on each side; the beam size 
fwhm is 17\arcsec. 
North is oriented upward and east to the left within each frame.
\label{images}}
\end{figure}

\end{document}